# NANOSCALE WEIBULL STATISTICS

N. M. Pugno[*] and R. S. Ruoff[#]

[*]Department of Structural Engineering, Politecnico di Torino,

Corso Duca degli Abruzzi 24, 10129, Italy

[#*]Department of Mechanical Engineering, Northwestern University,

Evanston, IL 60208-3111, USA

[#]r-ruoff@northwestern.edu; [*]nicola.pugno@polito.it

ABSTRACT

In this paper a modification of the classical Weibull Statistics is developed for nanoscale applications. It is called Nanoscale Weibull Statistics. A comparison between Nanoscale and classical Weibull Statistics applied to experimental results on fracture strength of carbon nanotubes clearly shows the effectiveness of the proposed modification. A Weibull's modulus of ~3 is, for the first time, deduced for nanotubes. The approach can treat (also) a small number of structural defects, as required for nearly defect free structures (e.g., nanotubes) as well as a quantized crack propagation (e.g., as a consequence of the discrete nature of matter), allowing to remove the paradoxes caused by the presence of stress-intensifications.

1. INTRODUCTION

Weibull statistics (Weibull, 1951) for strength (or time to failure, fatigue life, etc.) of solids and deterministic Linear Elastic Fracture Mechanics (LEFM; Griffith, 1920) do not apply properly at the nanoscale. Weibull statistics assumes that the number of critical flaws is proportional to the volume or to the surface area of the structure, whereas single crystal nanostructures are anticipated to be either defect-free or to have a small number of (critical) defects. Recently LEFM, which assumes infinite ideal strength of solids, as well as large (with respect to the so-called "plastic zone") and perfectly sharp cracks, has been modified and a new theory, Quantized Fracture Mechanics (QFM; Pugno and Ruoff, 2004), has been presented that quantizes the crack advancement. QFM is intended for treating defects of any size and shape (e.g., atomic vacancies, nano-holes). In this paper we present a modification of the Weibull statistics for describing the strength of solids (also) at the nanoscale. We apply this new statistical treatment to the largest collection of carbon nanotube strengths available (Yu et al., 2000). The Weibull modulus for nanotubes is obtained as ~3; furthermore, the statistical data analysis suggests that a small number of defects were critical for such nanotubes. An application to different types of whiskers is also discussed. The proposed approach, coupled with Quantized Fracture Mechanics, can treat not uniform stress distribution also if dominant stress-intensifications are present, thus removing the classical paradoxes related to the non convergence of the Weibull integrals.

## 2. CLASSICAL WEIBULL STATISTICS

Classical *Weibull Statistics* (Weibull, 1951) assumes the probability of failure $P_f$ for a specimen of volume $V$ under uniaxial stress $\sigma(V)$ as:

$$P_f(\sigma) = 1 - \exp\left[-\int_V \left(\frac{\sigma(V)}{\sigma_{0V}}\right)^m dV\right] \qquad (1a)$$

or equivalently:

$$P_f(\sigma) = 1 - \exp\left[-V^*\left(\frac{\sigma}{\sigma_{0V}}\right)^m\right] \qquad (1b)$$

where $\sigma_{0V}$ and $m$ are Weibull's scale (with anomalous physical dimension) and shape (dimensionless) parameters respectively, and $V^*$ is an "equivalent" volume that refers to a reference (e.g., the maximum) stress $\sigma$ in the specimen, defined by comparing eqs. (1a) and (1b) (see Bagdahn and Sharpe, 2003). If the specimen is under uniform tension $\sigma(V) \equiv \sigma$ and $V^* \equiv V$.

The surface-flaw based Weibull distribution simply replaces the volume $V$ in eqs. (1) with the surface area $S$ of the specimen (and $\sigma_{0V}$ with a new constant $\sigma_{0S}$):

$$P_f(\sigma) = 1 - \exp\left[-\int_S \left(\frac{\sigma(S)}{\sigma_{0S}}\right)^m dS\right] \qquad (2a)$$

$$P_f(\sigma) = 1 - \exp\left[-S^*\left(\frac{\sigma}{\sigma_{0S}}\right)^m\right] \qquad (2b)$$

The cumulative probability $P_f(\sigma_i)$ can be obtained experimentally as (Johnson, 1983):

$$P_f(\sigma_i) = \frac{i - 1/2}{N} \qquad (3)$$

where $N$ is the total number of tests and the observed strengths $\sigma_1,...,\sigma_N$ are ranked in ascending order.

The volume- and surface-based approaches become identical for the case of fracture of the external wall of nanotubes under (nearly) uniform tension, such as for the 19 nanotubes experimentally investigated by Yu et al. (2000, Table 1). This is true because $V = St = \pi DLt$, where $t$ is the constant spacing between nanotube walls (~0.34nm) and thus assigned as the shell thickness, and $D$ and $L$ are the nanotube diameter and length, respectively ($V^* \equiv V, S^* \equiv S$). The standard Weibull statistics applied to this set of fracture strength data is shown in Figure 1. The Weibull modulus is found to be ~3. This represents, according to our knowledge, the first estimation of the Weibull modulus for nanotubes. However, the correlation is very poor, showing a coefficient of correlation $R^2 = 0.67$. Perhaps such a statistics does not describe the real nature of strength of materials at the nanoscale.

3. NANOSCALE WEIBULL STATISTICS

According to QFM (Pugno and Ruoff, 2004) a quantized crack propagation has to be considered. QFM yields a better understanding of the experimental results and agrees with numerical simulations based on molecular mechanics and "ab initio" quantum mechanics (Mielke et al., 2004). The existence of a fracture quantum suggests that just a very small defect can cause the failure of a nearly defect free structure. For example, a single atomic vacancy (a very small hole) in an infinitely large graphene sheet reduces its strength by ~20% from the ideal strength (Pugno and Ruoff, 2004). Thus, at nanoscale just few defects can be responsible for the failure of the specimen, regardless its volume or surface. In addition, the tensional analog of the energy based QFM suggests that not the stress $\sigma$ but its mean value $\sigma^*$ along a fracture quantum has to reach a critical value to cause the failure of the specimen. Note that replacing $\sigma$ with $\sigma^*$ in the Weibull approach is sufficient to remove the classical paradoxes associated to the non convergence of the Weibull integrals at stress-intensifications (where the integral of $\sigma^m$ diverges whereas the integral of $\sigma^{*m}$ is finite).

Correspondingly, taking into account directly the number $n$ of critical defects and the quantized stress $\sigma^*$, from eqs. (1) and (2) we can formulate the *Nanoscale Weibull Statistics* (*NWS*) as:

$$P_f(\sigma^*) = 1 - \exp\left[-\sum_n \left(\frac{\sigma^*(n)}{\sigma_0}\right)^m\right] \tag{4a}$$

$$P_f(\sigma^*) = 1 - \exp\left[-n^*\left(\frac{\sigma^*}{\sigma_0}\right)^m\right] \tag{4b}$$

where $n^*$ is defined by comparing eq. (4a) and (4b) and can be considered an "equivalent" number of defects.

As an example we apply *NWS* to the experimental results on fracture strength of nanotubes by Yu et al. (2000). As previously described, the application of the Weibull statistics (identical for surface- or volume-based defects, as a consequence of the two dimensional nature of the experimentally stretched external nanotube walls) is shown in Figure 1.

The nanotubes were basically in uniform tension, thus $\sigma^*(n) \equiv \sigma^* \equiv \sigma$ and $n^* \equiv n$, where $\sigma$ is the applied load and $n$ is the number of critical defects. By applying *NWS* simply considering $n=1$, we find $m \sim 2.7$ (and $\sigma_0 \approx 31\text{GPa}$, see Figure 2) with a significantly better correlation of $R^2 = 0.93$ with respect to the interpretation based on the classical Weibull statistics (please also compare Figures 1 and 2).

# 4. COMPARISON BETWEEN CLASSICAL AND NANOSCALE WEIBULL STATISTICS

Let us assume fibers with circular cross section area (e.g., nanotubes) under uniform tension, i.e., $\sigma^*(n) \equiv \sigma^* \equiv \sigma$ and $n^* \equiv n$. The Weibull statistics assumes $n = kD^\alpha L^\beta$, with $\alpha = 2$ and $\beta = 1$ if volume-flaws are considered, or $\alpha = 1$ and $\beta = 1$ if surface-flaws are considered (and $k$ is a constant). On the other hand, we have noted that for nearly defect free structures, one may assume "point-flaws" defects, i.e., that failure occurs at $n=1$ (or equivalently at a value of $n$ independent from the specimen size) for which $\alpha = 0$, $\beta = 0$, so that in general, it may be more appropriate to expect $0 \leq \alpha \leq 2$ and $0 \leq \beta \leq 1$. For example, if "length-flaws" defects are considered $\alpha = 0$ and $\beta = 1$, i.e., $n \propto L$; for example for the nanotubes previously investigated this assumption would lead $m \sim 2.7$ and $R^2 = 0.74$. Thus, in our hypotheses, *NWS* considers $n = kD^\alpha L^\beta$ with $0 \leq \alpha \leq 2$, $0 \leq \beta \leq 1$ (or $n = kH^\alpha L^\beta W^\gamma$ for rectangular cross section areas $W \times H$, with $0 \leq \alpha, \beta, \gamma \leq 1$, e.g., nanowires). Accordingly, it is clear that *NWS* can be applied not only -but also- at nanoscale. We note that for such an example eq. (4) would correspond, for the limiting case of $\beta = 1$, to the modified Weibull distribution proposed by Zhu et al. (1997) in the study of the strength of sapphire whiskers and Nicalon SiC fibers. They showed that such a statistics includes all the three effects that have to be incorporated, according to Batdorf (1978), for a correct description of the strength of solids: (i) extreme value statistics (Gumbel, 1958), (ii) fracture mechanics (Griffith, 1920) and (iii) material characterization (e.g., dependence between length of the critical defect and specimen geometry). Thus, evidently, such effects are also included in our generalization, in which fracture mechanics is replaced by QFM.

Defining the nominal strength $\sigma_N$ of the material for a specified value of $P_f$, e.g., $P_f(\sigma = \sigma_N) = (1-e^{-1}) = 0.63$ ($\sigma_N$ is thus defined as the strength corresponding to the 63% probability of failure; $n = kD^\alpha L^\beta$) the corresponding size/shape-effect is predicted according to eq. (4) as:

$$\sigma_N = \sigma_0 k^{-1/m} D^{-\alpha/m} L^{-\beta/m} \tag{5}$$

Strictly speaking eq. (4) is defined for $\sigma < \sigma_C$ (here $\sigma^* \equiv \sigma$), where $\sigma_C$ is the (finite) ideal strength of solids, whereas obviously $P_f(\sigma \geq \sigma_C) \equiv 1$. Accordingly, in eq. (5) $\sigma_N$ is limited by $\sigma_C$. We note that the size-effect (thus assuming self-similar structures, i.e., $D \propto L$) predicted by eq. (5) is a power-law, in agreement with the fractal size effect law proposed by Carpinteri (1994 a,b; for a unified approach see also Carpinteri and Pugno, 2004). Note that the ratio between the exponents of $D$ and $L$ is equal to $\alpha/\beta$. In the classical Weibull statistics this ratio is set equal to 2 (volume-flaws) or 1 (surface-flaws). As recently emphasized by Zhu et al. (1997), the ratio $\alpha/\beta$ was observed to be significantly different for the sapphire ($\alpha - Al_2O_3$) whiskers studied by Bayer and Cooper (1967a). These whiskers were chemically polished to remove surface flaws, so that according to Weibull $\alpha/\beta \approx 1$ was expected. On the other hand, such a ratio was observed as even larger than 2 (that corresponds to volume-flaws): 7.0 for A-type

(fiber axis orientation $<11\bar{2}0>$ and $<10\bar{1}0>$, $\sigma_N \propto D^{-0.21}L^{-0.03}$), again 7.0 for C-type (axis orientation $<0001>$, $\sigma_N \propto D^{-0.14}L^{-0.02}$) or 15.4 for A-C-type (axis orientation $<10\bar{1}1>$, $\sigma_N \propto D^{-2.47}L^{-0.16}$). Furthermore, only for unpolished A-type sapphire whiskers did Bayer and Cooper (1967b) observe $\alpha/\beta \sim 1.43$ ($\sigma_N \propto D^{-0.56}L^{-0.39}$), thus in the range expected by the Weibull statistics. For unpolished C-type they observed no length dependence at all, and $\sigma_N \propto D^{-0.64}$. A similar strength dependence, as $\sigma_N \propto D^{-1}$, was observed in iron or copper whiskers by Brenner (1965). Thus, it is clear that such size/shape effects cannot be explained by Weibull statistics, whereas eq. (5) is compatible with the observations reported in the whisker literature (see also Levitt, 1970), as emphasized by Zhu et al. (1997) to demonstrate on sapphire whiskers the effectiveness of their Weibull modification (limit case of *NWS* for $\sigma^*(n) \equiv \sigma^* \equiv \sigma$ and $n^* \equiv n = kD^\alpha L^\beta$ with $\beta = 1$).

As a final example, we consider the $\alpha - Si_3N_4$ whiskers investigated by Iwanaga and Kawai (1998); they observed a maximum value of the strength equal to 59GPa (evidently close to the expected ideal material strength, see the first principles calculations by Ogata et al., 2004). A linear dependence for the whisker $\alpha - Si_3N_4$ strengths on their diameter was clearly observed (the whisker lengths were approximately constant and around 1-2 mm). We first assume the volume-flaw based Weibull statistics; fitting their data yields $m \sim 3.3$ ($R^2 = 0.89$) and $\sigma \propto D^{-0.61}$. Assuming surface flaws we find $m \sim 2.9$ ($R^2 = 0.89$) and $\sigma \propto D^{-0.34}$. Even if the observed dependence between strengths and diameters suggest that here considering *n*=1 is not realistic, since it would correspond to a size-independent strength (and for analogy to the volume- or surface-defects to "point-defects") such a case would correspond to $m \sim 2.5$ ($R^2 = 0.88$). Furthermore, fitting their experimental results on size- effects, we find $\sigma \propto D^{-0.4}$, suggesting that these failures were surface dominated. The example shows that for larger structures in general $n = kD^{\alpha}L^{\beta}$ has to be consider in the *NWS* rather than simply $n = 1$ (we note that the availability of only 6 strength values means that one should be cautious in "over-interpreting" the statistical fits).

5. CONCLUSIONS

The comparison between classical and Nanoscale Weibull Statistics applied to nanotubes clearly shows the effectiveness the proposed modification (also) for nanoscale applications. The Weibull's modulus for nanotubes is deduced as ~3. Comparing classic and nanoscale Weibull statistics it is also clear the role of the fracture quantization: this is crucial to treat stress-intensifications in the specimen, for which the classical Weibull integrals do not converge, in contrast to what happens in our treatment. Finally, the nanoscale statistical data analysis suggests that a small number of defects, perhaps simply one critical defect in each of the 19 different carbon nanotubes that were fractured, was responsible for breaking of these nanotubes.

**Acknowledgements**

The authors would like to thank A. L. Ruoff for critically reading and commenting the manuscript and A. Carpinteri for the helpful scientific discussions. RSR appreciates support from the NSF grant no. 0200797 "Mechanics of Nanoropes" (Ken Chong and Oscar Dillon, program managers), from the ONR grant no. N00014-02-1-0870 "Mechanics of Nanostructures" (Mark Spector and John Pazik, program managers), and from the NASA University Research, Engineering and Technology Institute on Bio Inspired Materials (BIMat) under award No. NCC-1-02037.

TABLE CAPTIONS

Table 1: Experimental results (Yu et al., 2000) on strength of multiwalled carbon nanotubes (only the external wall was fractured) and nanotube outer diameters and lengths.

FIGURE CAPTIONS

Figure 1: Weibull statistics for strength of carbon nanotubes (Table 1).

Figure 2: Nanoscale Weibull statistics for strength of carbon nanotubes (Table 1; $\sigma^*(n) \equiv \sigma^* \equiv \sigma_{applied}$ and $n^* \equiv n = 1$).

TABLES

| Test number | Diameter [nm] | Length [μm] | Strength [GPa] |
|---|---|---|---|
| 1 | 28.0 | 4.10 | 11 |
| 2 | 28.0 | 6.40 | 12 |
| 3 | 19.0 | 3.03 | 18 |
| 4 | 31.0 | 1.10 | 18 |
| 5 | 28.0 | 5.70 | 19 |
| 6 | 19.0 | 6.50 | 20 |
| 7 | 18.5 | 4.61 | 20 |
| 8 | 33.0 | 10.99 | 21 |
| 9 | 28.0 | 3.60 | 24 |
| 10 | 36.0 | 1.80 | 24 |
| 11 | 29.0 | 5.70 | 26 |
| 12 | 13.0 | 2.92 | 28 |
| 13 | 40.0 | 3.50 | 34 |
| 14 | 22.0 | 6.67 | 35 |
| 15 | 24.0 | 1.04 | 37 |
| 16 | 24.0 | 2.33 | 37 |
| 17 | 22.0 | 6.04 | 39 |
| 18 | 20.0 | 8.20 | 43 |
| 19 | 20.0 | 6.87 | 63 |

Table 1

FIGURES

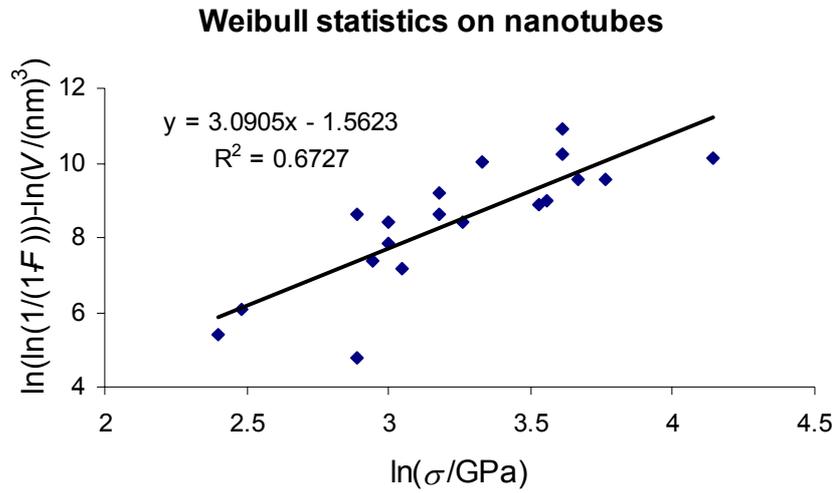

Figure 1

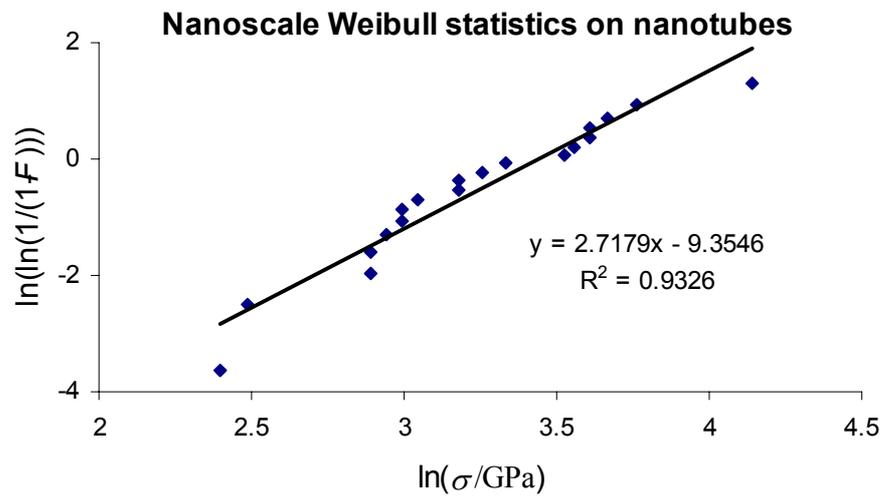

Figure 2